\documentstyle[12pt]{article}
\begin{document}
\begin{titlepage}
\begin{flushright}
Zurich University ZU-TH 27/96\\
\end{flushright}
\vskip 3cm
\begin{center}
{\large\bf 
BARYONIC DARK MATTER
\footnote{Talk presented by F. De Paolis at the workshop on
``Dark and visible matter in galaxies and cosmological implications'' 
(Sesto Pusteria, July 1996)}}
\vskip 1cm
F.~De Paolis$^{2,3}$,
G.~Ingrosso$^{2}$ ,
Ph.~Jetzer$^{3}$
and M.~Roncadelli$^{4}$
\vskip 0.5cm
$^2$
Dipartimento di Fisica, Universit\`a di Lecce, CP 193, 73100 Lecce, Italy\\
and INFN, Sezione di Lecce, CP 193, 73100 Lecce, Italy\\
$^3$ Paul Scherrer Institute, Laboratory for Astrophysics, \\
CH-5232 Villigen PSI,\\
and Institute of Theoretical Physics, University of Zurich, \\
Winterthurerstrasse 190, CH-8057 Zurich, Switzerland\\
$^4$ INFN, Sezione di Pavia, Via Bassi 6, I-27100, Pavia

\vskip 3cm
\end{center} 
\begin{abstract}
Reasons supporting the idea that most of the dark matter in 
galaxies and clusters of galaxies is baryonic are discussed. 
Moreover it is argued that most of the dark matter in galactic halos 
should be in the form of MACHOs and cold molecular clouds.
\end{abstract}

\end{titlepage}

\section{Introduction} 
One of the most important problems in modern astrophysics concerns the 
nature of the dark matter that 
pervades the Universe. Probably, more 
than $90\%$ of the matter in our Universe is dark.
Evidence for the existence of dark matter comes from the observation that 
the dynamics of many astronomical systems, such as galaxies and clusters of 
galaxies, cannot be explained by the visible matter alone. 

The most impressive evidence for dark matter is provided by measurements of 
the rotational velocities of stars and gas clouds  
in spiral galaxies. The dependence of the rotational speed from the 
galactocentric distance is, in fact, a measure of the total mass density 
profile $\rho(r)$. An important feature of many spirals is that 
their rotation curves, after an initial rise, remain almost constant 
with increasing galactocentric distance (Rubin, Ford \& Thonnard, 1980).
This fact yields convincing evidence that the dynamical mass of spirals 
increases much more rapidly than the visible matter with increasing 
radial distance from the galactic centre (see e.g . Persic, Salucci 
\& Stel 1996). 
Observations as well as 
theoretical arguments (linked to the stability of disks) suggest  that 
the dark matter is distributed in the 
form of spherical (or slightly oblate) halos around galactic disks. 

The presence of dark matter in elliptical galaxies is less evident as 
compared with the case of spirals, mainly due to the lack of neutral gas 
observable in the radio band (Sarazin 1987, Kent 1990).
Since the motion of 
stars in ellipticals is irregular and any rotation  
appears to be small, the Doppler-shift technique cannot be applied in 
the same way as for spirals. However, certain ellipticals are surrounded by 
a hot, ionized gas emitting in the X-ray band (due to Bremsstrahlung 
emission) which can be used to determine the galaxy gravitational potential.
Assuming that the gas is in hydrostatic equilibrium, the virial theorem 
relates the kinetic energy of the gas to its gravitational binding energy. 
It is then possible to show that, for the gas to remain confined within the 
galaxy, there must be much more mass than just the visible matter 
(Fabricant \& Gorenstein 1983, Fabricant, Lecar \& Gorenstein 1980, De Paolis,
Ingrosso \& Strafella 1995).

As far as clusters of galaxies are concerned, very convincing evidence for 
dark matter comes from investigations of the dynamical properties of galaxies, 
the gravitational lensing of distant background objects and the X-ray data 
from the Bremsstrahlung emission of fast moving electrons in the 
hot intergalactic plasma.

Candidates for dark matter can be divided into two categories: those 
motivated by particle physics and those suggested by astronomy. The first 
category includes weakly interacting massive particles (WIMPs): massive 
neutrinos, axions and particles predicted by supersymmetric theories (SUSY).
Among the particles candidates the neutrino is the only one known to exist. 
Many experiments have searched for neutrino masses but, so far, strong 
enough limits have not been put. 
Neutrinos have been produced during the Big Bang 
and contribute to the density of the Universe with a fraction 
$\Omega\simeq (n_{\nu}/n_{\gamma})(m_{\nu}/25~{\rm eV})$, where 
$n_{\nu}$ and $n_{\gamma}$ are the  number density of neutrinos and photons, 
respectively. Since most of the Big Bang models predict that the relic 
abundance of neutrinos is comparable with that of photons, one obtains that 
neutrinos with a mass $m_{\nu}\ge 25$ eV can close the Universe.
Candidates motivated by astronomy include brown dwarfs,
white dwarfs, neutron stars, black 
holes and cold clouds.

In the following sections we discuss several reasons that lead us to 
believe that 
most of the dark matter in galaxies and clusters of galaxies 
should be baryonic. Obviously, galaxy formation remains an open problem in 
this view, and the only explanation to date requires non-baryonic dark 
matter. Still, the point we want to make is that many properties of 
galaxies and clusters of galaxies are naturally accounted for by baryonic 
dark matter alone.

\section{MACHOs and molecular clouds}
From the standard Big Bang nucleosynthesis model (Copi, Schramm 
\& Turner 1995) one infers that $0.01<\Omega_B <0.1$ (Particle Data Group 
1996).
Since for the amount of luminous baryons 
one finds $\Omega_{\rm lum}\ll \Omega_B$, it follows that an important 
fraction of baryons are dark and they may well make  up the entire dark 
matter in galactic halos. 
The halo dark matter of our galaxy cannot be in the form of hot 
diffuse gas otherwise there would be a large X-ray flux, for which 
stringent upper limits are available. 
The abundance of neutral hydrogen gas is 
inferred from the 21-cm measurements, which show that its contribution is 
small. Another possibility is that the hydrogen gas is in molecular form 
clumped into cold clouds. Baryons could otherwise have been processed in 
stellar remnants. A natural option is provided by brown dwarfs with mass 
below $\sim 0.1~M_{\odot}$, which would be too light to ignite the hydrogen 
burning reactions. In principle, also M-dwarfs and white dwarfs could be 
conceived as dark matter candidates. However, a deeper analysis shows that 
the M-dwarf option is problematic. Indeed, optical imaging of high-latitude 
fields taken  with the Wide Field Camera of the Hubble Space 
Telescope indicates that less than a few percents of the galactic halo can 
be in this form (Bahcall et al. 1994, Hu et al. 1994). 
A scenario with white dwarfs as a major constituent of the galactic halo 
has been explored (see e.g. Adams and Laughlin 1996) but it requires a 
rather {\it ad hoc} initial mass function peaked somewhere around 
$2-6~M_{\odot}$. \footnote{
Moreover, a halo primarily made of white dwarfs would have left too much diffuse hot 
gas at temperature $\sim 2\times 10^6$ K emitting in the X-ray band.
Here we note that Adams and Laughlin (1996) do not take into account 
in their calculations the different 
evolution of low metallicity stars (that would produce the halo white 
dwarfs) with respect to stars with solar metallicity. White dwarfs in the 
galactic disk have average mass $\sim 0.7~M_{\odot}$, whereas 
white dwarfs originating 
from low metallicity stars are expected to have larger masses. Roughly 
speaking, the white dwarf mass is correlated to the mass of the star 
helium core, i.e. the mass inside the region where the radiative 
gradient $\nabla_{\rm rad}$ is larger than the adiabatic gradient 
$\nabla_{\rm ad}$.
While $\nabla_{\rm ad}$ is approximately constant, $\nabla_{\rm rad}$ strongly 
depends on the metallicity; this would generate more massive $He$ cores 
with decreasing star metallicity. This behaviour is also confirmed by 
numerical simulations (Lattanzio 1991) of the evolution of the stellar 
parameters from the Main Sequence to the first thermal pulse.
Another problem with the analysis of Adams \& Laughlin is the 
choice of $\sim 8-10~M_{\odot}$ as 
the mass above which a star becomes a supernova (SNII); 
due to the low metallicity 
of primordial halo stars, however, also stars with masses well below $8~
M_{\odot}$ will end their life as SNII, thus producing neutron stars 
(Jura 1986).
Se, we are then left with the only possibility that MACHOs (Massive 
Astrophysical Compact Halo Objects) are brown 
dwarfs with mass $\le 0.1~M_{\odot}$.
}
On the other hand, a substantial component of neutron stars 
and black holes with mass 
$\ge 1~M_{\odot}$ is excluded due to the overproduction of heavy elements.

Thus, brown dwarfs and cold molecular clouds are probably the best 
candidates for dark matter in galaxies.  
Paczy\'nski (1986) proposed the idea of 
using gravitational micro-lensing to detect massive dark object in our 
galaxy by monitoring the brightness of stars in the Magellanic Clouds.
The light from a distant star should in fact be deflected by the 
gravitational field of a massive object close to the line of sight from 
the Earth to the star. For typical masses and distances of halo dark objects
(i.e. for deflection angles $\sim 10^{-6}$ arcsec), the star will appear 
brighter than in the absence of a deflector.

Recent observations of microlensing events (Alcock et al. 1993, Aubourg et 
al. 1993) towards the Large Magellanic Clouds (LMC) 
suggest that MACHOs provide a substantial amount of the halo dark matter.
Assuming a standard spherical halo model (in which the MACHO velocity 
distribution function is taken to be Boltzmannian), it has been found 
that the 8 microlensing events found so far (Alcock et al. 1996)
imply a halo MACHO fraction as high as 50\% and an average mass 
of $0.27 ~M_{\odot}$ (Jetzer 1996). However, we note that the statistics of 
these events is at present too low to infer any definite
conclusion since both 
the halo fraction in the form of MACHOs and their average mass strongly 
depend on the assumed model for the visible and dark matter components of 
the galaxy (see De Paolis, Ingrosso \& Jetzer 1996).

Several authors have studied the problem of determining the
number of the expected microlensing events or, equivalently, 
the optical depth to microlensing by considering different
models for the mass distribution, both luminous and dark
in the galaxy (Kerins 1995, Kan-ya, Nischi \& Nakamura,
1995, Evans \& Jijina 1994, Evans 1994).
Recent upgraded self-consistent galactic models which include anisotropies 
in phase space for the MACHO distribution and a more realistic model for 
the distribution of the galactic luminous matter 
(De Paolis, Ingrosso \& Jetzer 1996) show that the 
values for the microlensing rate can decrease with respect 
to the standard values, thereby making brown dwarfs a plausible candidate 
for MACHOs. 

The problem arises of how MACHOs formed and in what form the remaining 
fraction of the galactic dark matter is. A scenario in which 
dark clusters of MACHOs and cold molecular clouds naturally form in 
the halo at large galactocentric distances has been recently proposed 
(De Paolis et al. 1995a-c) 
\footnote{Similar ideas have been put forward by Gerhard \& Silk (1995).} 
and several methods to test this model have been 
proposed (De Paolis et al. 1995d). Basically, here the dynamics 
of the formation of dark 
clusters is similar to that of stellar globular clusters, the only 
difference being the larger galactocentric distance of dark clusters and 
consequently the lower incoming UV flux (from a central source).
This fact implies that molecular hydrogen in dark clusters is not 
dissociated so that the Jeans mass can naturally reach values as low as
$\sim 10^{-2}-10^{-1}~M_{\odot}$, leading to the formation of MACHOs.
It is important to note that also molecular clouds should form in dark 
clusters, since the process leading to MACHO formation does not have a 
$100\%$ efficiency  and the gas cannot be expelled due to 
the absence of strong stellar winds.

\section{Dark matter  at the centre of galaxies}
Recent observations of the central flatness of the velocity profiles of dark 
halos  seem to suggest that dark matter in galaxies is baryonic.
In fact, several computer simulations of the large scale structure of the 
Universe with non-baryonic dark matter (and in particular with {\it cold} 
vvdark matter) with a sufficiently high resolution 
to resolve the internal structure of the galactic halos, seem to 
indicate that the density profiles should have central cusps. 
These cusps are incompatible with the isothermal density profiles
\begin{equation}
\rho(r)=\frac{\rho(0)}{1+(r/a)^2}~, \label{1}
\end{equation}
where $a$ is the dark matter core radius. While this profile becomes 
approximately constant at $r\ll a$ and has a finite central density 
$\rho(0)$, numerical simulations indicate a density distribution that diverges 
like $r^{-1}$ (Burkert 1995). 
The existence of a central density cusp in 
normal galaxies is difficult to demonstrate since the internal regions 
are gravitationally dominated by the visible component. The 
distribution of dark matter, in fact, strongly depends on both the assumed
mass/luminosity ratio (M/L) for the disk and for the central spheroidal 
component. 
The situation is different in dwarf galaxies which have recently been
studied using high resolution methods. 
Dwarf spiral galaxies provide excellent probes for the internal 
structure of dark halos since these galaxies are completely dominated by dark 
matter on scales larger than a kiloparsec (Carignan \& Beaulieu 1989, 
Lake, Schommer \& van Gorkom 1990, Jobin \& Carignan 1990, Carignan \& 
Freeman 1988).
One can, therefore, use these galaxies to investigate the inner structure of 
dark halos with very little ambiguity about the contribution from the 
luminous matter and the resulting uncertainties in the disk M/L ratio.
Only about a dozen rotation curves of dwarf galaxies have been measured, 
but a trend clearly emerges: the rotational velocities rise over most of the 
observed region, which spans several times the optical scale lengths and
nevertheless lies within the core radius of the mass distribution.
Rotation curves of dwarf galaxies  do not admit singular density 
profiles at the galactic centre and their profiles are in good agreement 
(see Flores \& Primack 1988, Moore 1994)
with Eq. (\ref{1}).
The shape of the central cores of dark 
matter can be explained with the model we have briefly discussed in Sect. 2 
in which the dark matter in galaxies is 
constituted by dark clusters (DC) of MACHOs and cold molecular clouds that 
should form mainly at distances larger than $R_{\rm crit}\sim 5-10$ 
kpc. Baryonic dark matter inside $R_{\rm crit}$ should derive from 
DCs broken during the galactic evolution (due to encounters among DCs or 
passages of them through the disk). 
This gas should form stars or remain in gaseous form, but the dark 
matter can never have a central density cusp since the gas has to thermalize 
and then its density profile is still dictated by eq. (\ref{1}).

\section{Galactic evolution along the Hubble sequence}
It has been recently pointed out (Pfenniger, Combes \& Martinet 1994)
that spiral galaxies 
evolve along the Hubble sequence from $S_d$ to $S_a$ in  billions of 
years. During this evolution the dimensions of both galactic nuclei 
and disks increase while the M/L ratio should decrease. This fact
suggests that dark matter gradually transforms into visible matter, 
that is in stars. 
Of course, this is possible only if the dark matter is baryonic and in particular if 
it is in gaseous form.
In the scenario of Pfenniger, Combes and Martinet (1994) the dark matter, 
in the form of self-gravitating $H_2$ clouds, is in the galactic disk.
The clouds have a fractal structure that 
ranges upwards over 4 to 6 orders of magnitude in scale.
The elementary cloudlets have low temperature $\sim 3$ K, typical 
number density
$\sim 10^9~ {\rm cm}^{-3}$, size $\sim 5 \times 10^{-6}$ pc 
and mass $\sim 10^{-3}~M_{\odot}$. 
On the contrary, in the framework of our model (De Paolis et al. 1995a-d e 
1996a),
the galactic dark matter is composed of MACHOs and molecular clouds located 
in the galactic halo.

\section{Rotation curve shapes}
Initial studies have indicated that
rotation curves of spiral galaxies are generally flat. 
This means that the galactic halo 
must produce practically the entire rotational velocity far out the optical 
radius, while in the internal regions the optical disk maximally 
contributes to the rotation curve. It seems that disk and halo 
combine together to produce a flat rotation curve. This synthony between 
disk and halo has been called the {\it disk-halo conspiracy} 
(van Albada \& Sancisi 1986, Sancisi \&  van Albada 1987).
However, more recent observational results indicate that this 
{\it conspiracy} is not always true. 
For example, in some dwarf irregular galaxies the dark 
halo mass is considerably higher than the luminous disk mass inside the 
optical radius. Persic and Salucci (1988) developed a method to estimate 
the dark matter fraction inside $R_{25}$ by using the shape 
of the rotation 
curves in the optical band. This study suggests that the fraction of the 
dark matter increases as the luminosity of spiral galaxies decreases. Then, 
in the internal regions of bright spirals the disk is the dominant 
component while the halo contributes significantly to the rotation curve 
only at large distances from the galactic centre. Vice versa, in low 
luminous galaxies, the dark halo seems to dominate at all scales. 
Based on these results one can predict that rotation curves in bright spirals 
should decrease at distances larger than the optical radius while in low 
luminous galaxies the rotation curve has to remain flat or increase 
beyond the optical radius.
Combining the rotation curves obtained from HI observations for 
many spirals with the curves reported in the literature (Casertano \& 
Van Gorkom 1991), the decrease 
of the rotational velocity beyond the optical radius in bright spirals has 
been identified. As predicted, the rotational
velocities turn out to be flat or to increase in low luminous spiral galaxies.
Moreover, a correlation between  the maximum velocity 
and the slope of the rotation curve beyond the optical radius has been 
found.
The dependence of the rotation curves on the luminous content of the spiral 
galaxies we are talking about can be explained if the dark matter in 
spirals is baryonic and in 
particular if halos formed before galactic disks (Ashman 1992, Persic \& 
Salucci 1991), as it naturally happens in our model (De Paolis et al. 1995a-d).

\section{Dark matter in clusters of galaxies}
It is well known (see e.g. Blumenthal et al. 1984) that the 
ratio M/L increases from the luminous part of galaxies to clusters 
and superclusters of galaxies. This fact has 
generally induced astrophysicists to conclude that clusters and 
superclusters of galaxies have much more matter per unit luminous matter 
than individual galaxies, so that the critical density of the Universe 
($\Omega=1$) can be attained.

Recently, it has been shown (Bahcall, Lubin \& Dorman 1995) that most of the 
dark matter in clusters and superclusters of galaxies should be clumped in the 
halos around galaxies. Indeed, the ratio M/L in clusters does not 
significantly increase at scales larger than 100-200 kpc, typical of 
galactic halos. The total mass of the clusters can then be accounted for 
by the mass of the individual galaxies plus the hot gas mass 
(that represents less than $\sim 20\%$ of the total cluster mass). 
This fact suggests that $\Omega$ can be as low as 
$\sim 0.2$, marginally consistent with the limits from
the Big Bang nucleosynthesis (Sect. 2).

\section{Conclusions}
The wide-spread opinion in the scientific community that dark matter in 
galaxies and clusters of galaxies cannot be baryonic looks as an unwarranted 
conclusion.
As we have discussed in the previous sections, there are many reasons 
to believe that dark matter in galaxies and clusters of galaxies is baryonic. 
The most likely dark matter candidates in galaxies are brown dwarfs and cold 
molecular clouds (Sect. 2). 

MACHOs have been discovered by gravitational microlensing events towards 
stars in the LMC. It is unlikely that MACHOs are hydrogen-burning stars 
with masses in the range $0.1-0.3~M_{\odot}$ due to their emission in the 
infrared band (Hu et al. 1994). 
Also the possibility that MACHOs are white dwarfs 
with masses of $0.2-0.3~M_{\odot}$ gives rise to various problems, 
as discussed in section 2. We note that in the proposed model for 
the galactic dark matter 
it can be expected that a considerable number of MACHOs is in binary 
systems. This is expecially true for MACHOs in the core of the 
dark clusters (see De Paolis et al. 1996b). This can obviously explain an 
increase up to a factor of 2 in the MACHO mean mass as measured by
microlensing experiments towards stars in the LMC.

We have discussed many evidences for baryonic dark matter. These evidences 
arise from the shape of the dark 
matter density profiles towards the centre of dwarf galaxies, from the 
evolution of spiral galaxies along the Hubble sequence, and from
the shape of rotation 
curves of spiral galaxies. On larger scales, another evidence 
supporting a baryonic scenario comes from the analysis of the M/L 
ratio of clusters of galaxies.

\section{References}

\noindent
Adams, F. C. \& Laughlin, G. 1996, astro-ph/9602006

\noindent
Alcock, C. {\it et al.} 1993, {\it Nature}, 365, 621

\noindent
Alcock, C. {\it et al.} 1996, astro-ph 9606165

\noindent
Ashman, K. M. 1992, PASP, 104, 1109

\noindent
Aubourg, E. {\it et al.} 1993, {\it Nature}, 365, 623

\noindent
Bahcall, J., Flynn, C., Gould, A \& Kirhakos, S. 1994, ApJ, 
435, L51

\noindent
Bahcall, N. A., Lubin,  L.M. \& Dorman, V. 1995, ApJ, 447, L81

\noindent
Blumenthal, G. R., Faber,  S. M., Primack, J. R. \& Rees, M. J.
1984, {\it Nature}, 311, 517

\noindent
Burkert, A. 1995, ApJ 447, L25 

\noindent
Carignan, C. \& Beaulieu, S. 1989, ApJ 347, 760 

\noindent
Carignan, C. \&  Freeman, K. C. 1988, ApJ 332, L33

\noindent
Casertano, S. \& Van Gorkom, J. H. 1991, ApJ 101, 1231 

\noindent
Copi, C. J., Schramm, D. N. \& Turner, M. S. 1995, {\it 
Science}, 267, 192

\noindent
De Paolis, F., Ingrosso, G. \& Jetzer, Ph. 1996, 
ApJ October 10, 1996 issue (in press)

\noindent
De Paolis, F., Ingrosso, G., Jetzer,  Ph. \& Roncadelli, M. 
1995a, {\it Phys. Rev. Lett.}, 74, 14

\noindent
De Paolis, F., Ingrosso, G., Jetzer,  Ph. \& Roncadelli, M. 
1995b, Astron. \& Astrophys. 295, 567

\noindent
De Paolis, F., Ingrosso, G., Jetzer,  Ph. \& Roncadelli, M. 
1995c,
{\it Comm. Astrophys.}, 18, 87

\noindent
De Paolis, F., Ingrosso, G., Jetzer, Ph., Qadir, A. \& Roncadelli, 
M. 1995d, Astron. \& Astrophys.  299, 647

\noindent
De Paolis, F., Ingrosso, G., Jetzer,  Ph. \& Roncadelli, M. 
1996a, {\it Int. J. Mod. Phys. D}, 5, 151

\noindent
De Paolis, F., Ingrosso, G., Jetzer,  Ph. \& Roncadelli, M. 
1996b, in preparation

\noindent
De Paolis, F.,  Ingrosso, G. \& Strafella, F. 1995, ApJ 438, 83

\noindent
Evans, N. W. \& Jijina, J. 1994, MNRAS 267, L21

\noindent
Evans, N. W. 1994, MNRAS 267, 333

\noindent
Fabricant, D. \& Gorenstein, P. 1983, ApJ 267, 535 

\noindent
Fabricant, D., Lecar, M. \& Gorenstein, P. 1980, ApJ 241, 552

\noindent
Flores, R. A. \&  Primack, J. R. 1988, ApJ 427, L1

\noindent
Gerhard, O. \& Silk, J. 1995, astro-ph 9509149

\noindent
Hu, E. M., Huang, J. S., Gilmore, G. \& Cowie, L. L. 1994, 
{\it Nature}, 371, 493

\noindent
Jetzer, Ph 1996, to appear in {\it Helv. Phys. Acta}

\noindent
Jobin, M. \& Carignan, C. 1990, Astron. J., 100, 648

\noindent
Jura, M. 1986, ApJ 301, 624

\noindent
Kan-ya, Y., Nischi, R. \& Nakamura, T. 1995, astro-ph 9505130

\noindent
Kent, S. M. 1990, in {\it Evolution of the Universe}, ASP 
Conference Series Vol.
10, Ed. R. G. Kron, Brigham Young Univ. Printing Services, pag. 109

\noindent
Kerins, E. 1995, MNRAS in press (astro-ph 9406040

\noindent
Lake, G., Schommer,  R. A. \& van Gorkom, J. H. 1990, Astron. J. 99, 547

\noindent
Lattanzio, J. C. 1991, ApJS 76, 215

\noindent
Moore, B. 1994, {\it Nature},  370, 629

\noindent
Paczy\'nski, B. 1986 ApJ 304, 1

\noindent
Particle Data Group 1996, {\it Phys. Rev.}, D54, 109

\noindent
Pfenniger, D., Combes, F. \& Martinet, L. 1994, Astron. \& Astrophys. 285, 79

\noindent
Persic, M. \& Salucci, P. 1988, MNRAS 234, 131

\noindent
Persic, M. \& Salucci, P. 1991, MNRAS 248, 325

\noindent
Persic, M., Salucci, P. \& Stel, F. 1996, MNRAS 281, 27

\noindent
Rubin, V. C., Ford, W. K. \& Thonnard, H. 1980, ApJ 238, 471 

\noindent
Sancisi, R. \&  van Albada,  T. S. 1987,
{\it IAU Symp. 117}, {\it Dark matter in the Universe}, 
Eds. J. Kormendy, G. R. 
Knapp, Reidel Publishing Company, pag. 67 

\noindent
Sarazin, C. L. 1987, in 
{\it Structure and Dynamics of Elliptical Galaxies}, IAU 
Symp. 127, Ed. P. T. de Zeuw, Reidel, Dordrecht, pag. 179

\noindent
van Albada, T. S. \& Sancisi, R. 1986, {\it Philos. Trans. 
Royal Soc. Lond. A}, 320, 447 

\end{document}